\documentclass[12pt]{iopart}

\usepackage{graphicx,amssymb,url}
\usepackage{wasysym}
\begin{document}

\title{Bridgeness: A Local Index on Edge Significance in Maintaining Global Connectivity}

\author{Xue-Qi Cheng$^{1,2}$\footnote{Author to whom any correspondence should be addressed.}, Fu-Xin Ren$^{1}$, Hua-Wei Shen$^{1}$, Zi-Ke Zhang$^{3}$, Tao Zhou$^{2,4}$}

\address{$^{1}$ Institute of Computing
Technology, Chinese Academy of Sciences, Beijing 100190, People's
Republic of China}
\address{$^{2}$ Web Sciences Center, University of
Electronic Science and Technology of China, Chengdu 610054, People's
Republic of China}
\address{$^{3}$ Department of Physics, University of
Fribourg, Chemin du Mus\'ee 3, CH-1700 Fribourg, Switzerland}
\address{$^{4}$ Department of Modern Physics, University of Science and
Technology of China, Hefei 230026, People's Republic of China}

\eads{\mailto{cxq@ict.ac.cn},
\mailto{renfuxin@software.ict.ac.cn},
\mailto{shenhuawei@software.ict.ac.cn},
\mailto{zhangzike@gmail.com},
\mailto{zhutou@ustc.edu}}

\date{\today}

\begin{abstract}
Edges in a network can be divided into two kinds according to their
different roles: some enhance the locality like the ones inside a
cluster while others contribute to the global connectivity like the
ones connecting two clusters. A recent study by Onnela \emph{et al} 
uncovered the weak ties effects in mobile communication. In this article, we
provide complementary results on document networks, that is, the
edges connecting less similar nodes in content are more significant
in maintaining the global connectivity. We propose an index named
\emph{bridgeness} to quantify the edge significance in maintaining
connectivity, which only depends on local information of network
topology. We compare the bridgeness with content similarity and some
other structural indices according to an edge percolation process.
Experimental results on document networks show that the bridgeness
outperforms content similarity in characterizing the edge
significance. Furthermore, extensive numerical results on disparate
networks indicate that the bridgeness is also better than some
well-known indices on edge significance, including the Jaccard
coefficient, degree product and betweenness centrality.

\end{abstract}


\maketitle

\section{Introduction}

Recently, the study of complex networks became a common focus in
many branches of science
\cite{Albert2002,Newman2003,Boccaletti2006,Newman2006}. Many
measurements are used to characterize the role of a node in network
structure and function, ranging from simple indices like degree and
closeness \cite{Sabidussi1966} to complicated centralities like
betweenness \cite{Freeman1977}, PageRank score \cite{Brin1998} and
some random-walk-based indices \cite{Liu2010}. In comparison, the
study of edge's role is less extensive. Actually, an edge may play
an important role in enhancing the locality or be significant in
maintaining the global connectivity. How to differentiate these two
roles is an interesting and unsolved problem. A related issue is the
famous \emph{weak ties theory} \cite{Granovetter1973}, which tells
us that the job opportunities and new ideas usually come from people
with weak connections. Further more, the weak ties can enhance the
spreading of rumor \cite{Lai2002} and knowledge \cite{Levin2004},
keep the stability of biological functions \cite{Csermely2004} and
improve the accuracy of network structure prediction \cite{Lu2010}.
Moreover, experimental and theoretical analysis on a large-scale mobile
communication network showed that the weak ties play the leading
role in maintaining the global connectivity \cite{Onnela2007},
and they are important in the process of the emergence of social
communities \cite{Kumpula2007}, which is a common topological characteristics
of networks and has been extensively studied in the literature of
network theory \cite{Girvan2002,Cheng2010,Shen2010}. In \cite{Onnela2007},
the strength of a tie is quantified by the time spent on
communication, while in this article we consider a similar problem
on document networks where the tie strength is characterized by
content similarity. Analogous to the observation of the social
communication network, we find that the ties connecting less similar
documents are more significant in maintaining global connectivity.

Generally speaking, the strength information, such as the
communication time and content similarity, is not easy to be
obtained. Therefore, to quantify the edge significance only making
use of the information of observed topology is very valuable in
practice. For this purpose, we propose the index \emph{bridgeness}.
To our surprise, this index can better characterize the edge
significance in maintaining the global connectivity than the content
similarity in document networks. Furthermore, extensive numerical
results on disparate networks indicate that the bridgeness is also
better than some well-known structural indices on edge significance,
including the \emph{Jaccard coefficient} \cite{Jaccard1901}, \emph{degree
product} \cite{Holme2002,Giuraniuc2005,Moreira2009,Hooyberghs2010} and \emph{edge betweenness} \cite{Girvan2002}.

\section{Weak Ties Phenomenon in Document Networks}

Social networks usually exhibit two important phenomena, being
respectively the homophily \cite{Lazarsfeld1954,McPherson2001} and weak ties effects
\cite{Granovetter1973}. Homophily tells us that connections are more
likely to be formed between individuals with close interests,
similar social statuses, common characteristics or shared
activities. The well-known transitivity or triadic closure
phenomenon \cite{Rapoport1953} can be taken as a reflection for homophily.
Homophily contributes to the formation of most connections in social
networks. In contrast, some connections are formed between less
similar individuals, which usually have weaker strength. Weak ties
phenomenon refers to the fact that these connections may play
crucial roles in maintaining the global connectivity and holding key
functions, such as the flow of new information and knowledge.

Compared with social networks, the homophily and weak ties
phenomenon are less studied for document networks. In document
networks, each node corresponds to a document with textual contents.
Typical examples are the World Wide Web and scientific citation
networks. Two documents are more likely to be connected if they are
relevant, e.g., webpages presenting the same topic and articles
belonging to the same research area. The probability that two
documents are connected increases with their content similarity
\cite{Menczer2004} and the probability that three documents form a
triangle increases with their trilateral content similarity
\cite{Cheng2007,Cheng2009}. These indicate that the content
similarity in document networks plays the similar role as the tie
strength in social networks.

Inspired by Onnela \emph{et al} \cite{Onnela2007}, we quantify the
weak ties phenomenon according to an edge percolation process.
Generally speaking, if the weak ties phenomenon exists in terms of
content similarity, the network will disintegrate much faster when
we remove edges successively in ascending order of content
similarity than in descending order. Two quantities are used to
describe the percolation process. The first one is the fraction of
nodes contained in the giant component, denoted by $R_{GC}$. A
sudden decline of $R_{GC}$ will be observed if the network
disintegrates after the deletion of a certain fraction of edges.
Another quantity is the so-called \emph{normalized susceptibility},
defined as
\begin{equation}
\tilde{S} = \sum_{s<s_{max}}\frac{n_s s^2}{N},
\end{equation}
where $n_s$ is the number of components with size $s$, $N$ is the
size of the whole network, and the sum runs over all components but
the largest one. Considering $\tilde{S}$ as a function of the
fraction of removed edges $f$, usually, an obvious peak can be
observed that corresponds to the precise point at which the network
disintegrates \cite{Stauffer1994,Bunde1996}.

\begin{table}\caption{\label{tab:Data_description}
Basic statistics of the four networks: the PNAS citation network,
social network extracted from del.icio.us, political blog network
and scientific collaboration network from astrophysicists. $N$ and
 $L$ denote the number of nodes and edges in the
original network, while $N_{CG}$ and $L_{CG}$ are the number of
nodes and edges in the giant component. } \footnotesize\rm
\renewcommand{\arraystretch}{1.3}
\begin{center}
\begin{tabular}[]{ccccc}
\br
Networks/Measures & $N$   & $L$ & $N_{GC}$  & $L_{GC}$ \\
\hline
PNAS & 28,828 & 40,393 & 21,132 & 38,723 \\
Delicious & 64,615 & 107,140 & 64,615 &  107,140\\
Political Blog& 1490 & 16,715 & 1222 &  16,714\\
Astro Collaboration & 16,706 & 121,251 & 14,845 & 119,652 \\
\br
\end{tabular}
\end{center}
\end{table}

We use the PNAS citation network to illustrate the weak ties
phenomenon, where each node represents a paper published in the
\emph{Proceedings of the National Academy of Sciences of the United
States of America} (PNAS, http://www.pnas.org), and undirected edges
represent the citations. The data set contains all the papers from
1998 to 2007, and the basic statistics of this network is shown in
Table~\ref{tab:Data_description}. We collect keywords presented in titles and abstracts of
papers and construct a keyword vector space according to the
standard procedures in information retrieval \cite{Salton1989}. Here
keywords refer to those words which are not stop words such as
``a'', ``the'', ``of'', etc. The content of a paper is then
represented as a keyword vector, $\overrightarrow{X}$, which gives
the frequency of each keyword. The content similarity $R$ between
two papers, $i$ and $j$, is thus defined as:
\begin{equation}
R_{ij}={R_{ji}}=\frac{\|\overrightarrow{X_i} \cdot
\overrightarrow{X_j}\|}{\|\overrightarrow{X_i}\|
\|\overrightarrow{X_j}\|}. \label{eq:cos}
\end{equation}

\begin{figure}
\begin{center}
\includegraphics[width=14cm]{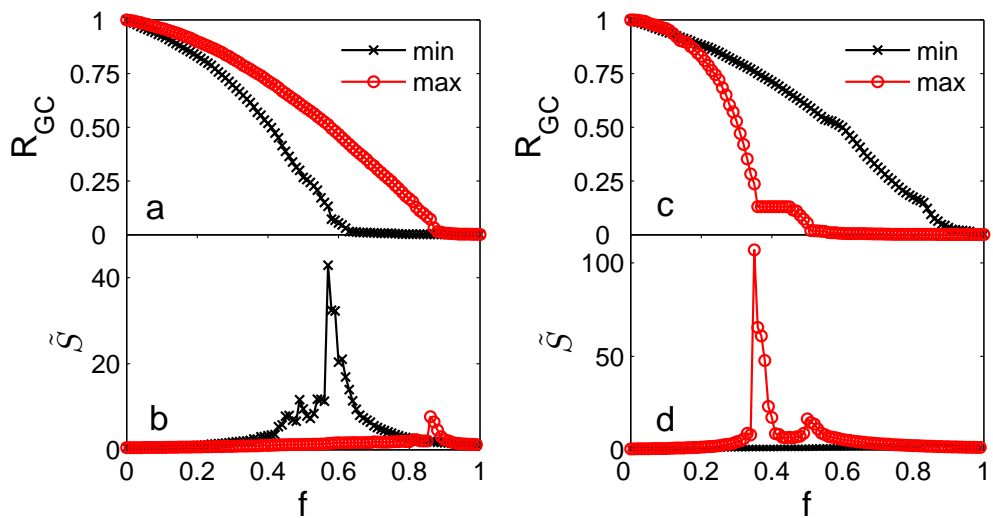}
\caption{\label{Link_Percolation_Result_PNAS} Edge
percolation results on PNAS citation network. Plots (a) and (b) are
for the content similarity, while (c) and (d) are for the
bridgeness. In (a) and (b), the $min$-($max$-)lines represent the
processes that the edges are removed from the least(most) similar to
the most(least) similar ones. In (c) and (d), the
$min$-($max$-)lines represent the processes that the edges with
smaller(larger) bridgeness are earlier to be removed.}
\end{center}
\end{figure}

Figure.~\ref{Link_Percolation_Result_PNAS} shows the edge
percolation results on the PNAS citation network. As shown in
Fig.~\ref{Link_Percolation_Result_PNAS}a, $R_{GC}$ decays much
faster when we firstly remove the less similar edges.
Correspondingly, in Fig.~\ref{Link_Percolation_Result_PNAS}b, a
sharp peak appears in the edge removing process from the weakest to
the strongest ones, indicating a percolation phase transition. While
if we firstly remove stronger edges, no clear peak can be observed.
Analogous to the report on mobile communication network
\cite{Onnela2007}, this result strongly supports the weak ties
phenomenon in the document network.

\section{Bridgeness vs. Tie Strength}

As mentioned above, tie strength is a good indicator for edge's role
of maintaining global connectivity. However, the strength
information is usually hard to be obtained. In social networks it
requires the personal information or historical activities, while in
document networks it has to crawl the content and calculate the
textual similarity. The processes are probably complicated, time
consuming, or even infeasible. Therefore, an index depending only on
topological information is of great advantage in practice. As we have
mentioned, in social and document networks similar or relevant nodes are
apt to connect to others and form local clusters. Clique is
the simplest structure to describe the locality in a network
\cite{Cheng2009}. A clique of size $k$ is a fully connected subgraph
with $k$ nodes \cite{Xiao2007}, and the clique size of a node $x$ or
an edge $E$ is defined as the size of maximum clique that contains
this node or this edge \cite{Shen2009,Shen20092}. We found that
in social and document networks we observed, the clique structures of
different sizes are prevalent. Recently, Gonz\'alez {\it et al} \cite{Gonzalez2007}
have reported the presence of k-clique communities in school friendship networks from the Add Health
database and the topology of c-network formed by communities. In his work
nodes shared by different communities are critical in connecting the c-network. Here we
focus on the roles of edges, i.e., edges in cliques mainly contribute to
locality while those between cliques are important in
connecting the network. Accordingly, we define the bridgeness of an edge as
\begin{equation}
B_E = \frac{\sqrt{S_x S_y}}{S_E}, \label{eq:Bridgeness_Definition}
\end{equation}
where $x$ and $y$ are the two endpoints of the edge $E$. For
example, in Fig. 2, the clique sizes of $x$, $y$ and $z$ are
$S_x=5$, $S_y=4$ and $S_z=3$, and the clique sizes of $E$, $E'$ and
$E''$ are $S_E=S_{E'}=S_{E''}=3$.

\begin{figure}
\begin{center}
\includegraphics[width=8cm]{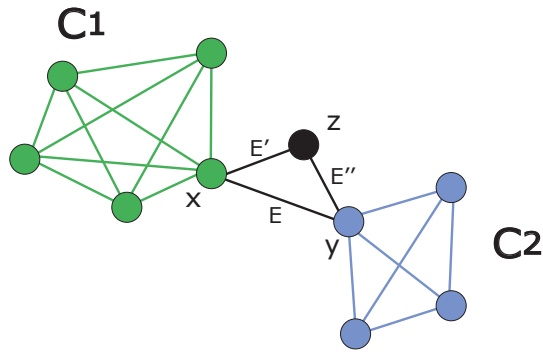}
\caption{\label{Bridgeness_Example} Illustration of
the bridgeness. $C_1$ and $C_2$ are two cliques of size 5 and 4, and
the edge $E$ connects them is of bridgeness 1.49.}
\end{center}
\end{figure}

\begin{figure}
\begin{center}
\includegraphics[width=9cm]{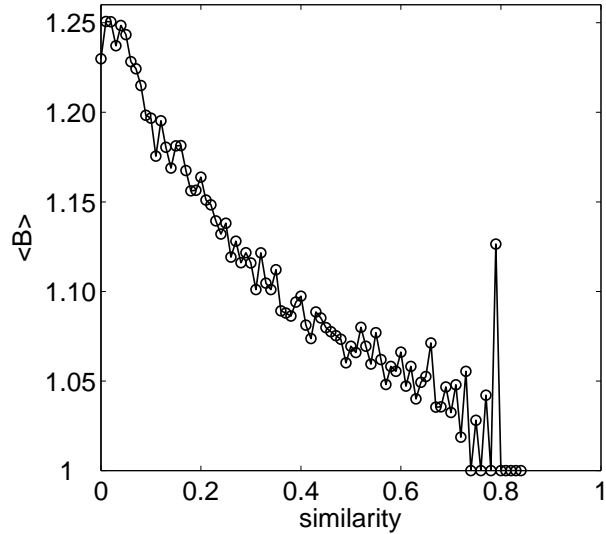}
\caption{\label{Bridgeness_Strength_Relation} The relation between
bridgeness and content similarity in the PNAS citation network.
$\left \langle B \right \rangle$ is the average bridgeness value of
edges with the same content similarity.}
\end{center}
\end{figure}

An interesting and surprising result can be observed from
Fig.~\ref{Bridgeness_Strength_Relation} that there is a negative
correlation between bridgeness and content similarity, namely the
weaker content similarity between two papers is, the stronger its
bridgeness will be. Combined with the conclusion in the previous
section, we infer that edges with strong bridgeness will play a more
important role in maintaining the global connectivity.
Fig.~\ref{Link_Percolation_Result_PNAS}c shows that if the edges
with larger bridgeness are removed first (corresponding to the curve
labeled by \emph{max}), the network quickly splits into many pieces,
while if we start removing the edges with smaller bridgeness
(corresponding to the curve labeled by \emph{min}), the network
gradually shrinks. As shown in
Fig.~\ref{Link_Percolation_Result_PNAS}d, a sharp peak is observed
in the former case. In addition, the critical point (corresponding
to the peak in $\tilde{S}(f)$) in Fig. 1d is remarkably smaller than
the one in Fig, 1b, indicating that the bridgeness can even better
characterize the edge's significance in maintaining the global
connectivity than the content similarity.

\section{Bridgeness vs. Other Indices}

In this section we compare the bridgeness with some other indices
for edge significance, which are also dependent only on the
topological structure. Three representative indices used for
comparison are introduced as follows (for more information about how
to quantify the edge significance, please see Refs.
\cite{Liben-Nowell2007,Fouss2007,Zhou2009}).
\begin{itemize}
  \item \emph{Jaccard coefficient} \cite{Jaccard1901}. Jaccard coefficient is defined as
  \begin{equation}
  J_E=\frac{|U_x \cap U_y|}{|U_x\cup U_y|},
  \end{equation}
where $x$ and $y$ are the two endpoints of the edge $E$ and $U_x$ is
the set of $x$'s neighboring nodes.
  \item \emph{degree product}. Degree product is defined as
  \begin{equation}
  D_E = k_xk_y,
  \end{equation}
where $k_x$ is the degree of node $x$. An extended form, $(k_xk_y)^\alpha$
has been recently applied in the studies of biased percolation
\cite{Giuraniuc2005,Moreira2009,Hooyberghs2010}. Notice that we only
care about the order of the edges, so the values of $\alpha$ (if $\alpha>0$)
play no role on the results and we there choose $\alpha=1$, the same as in \cite{Holme2002}.

  \item \emph{edge betweenness centrality} \cite{Girvan2002}.
Edge betweenness centrality counts for the number of shortest paths
between pairs of nodes passing through the edge, as
\begin{equation}
C_E = \sum_{s\neq t}\frac{{\sigma_{st}{(E)}}}{\sigma_{st}},
\end{equation}
where $\sigma_{st}$ is the number of shortest paths from node $s$ to
node $t$, and $\sigma_{st}{(E)}$ is the number of shortest paths
from $s$ to $t$ that pass through edge $E$. Notice that this index
is a global index whose computing complexity is remarkably higher
than the three local indices -- bridgeness, Jaccard coefficient and
degree product.
\end{itemize}

\begin{figure}
\begin{center}
\includegraphics[width=17cm]{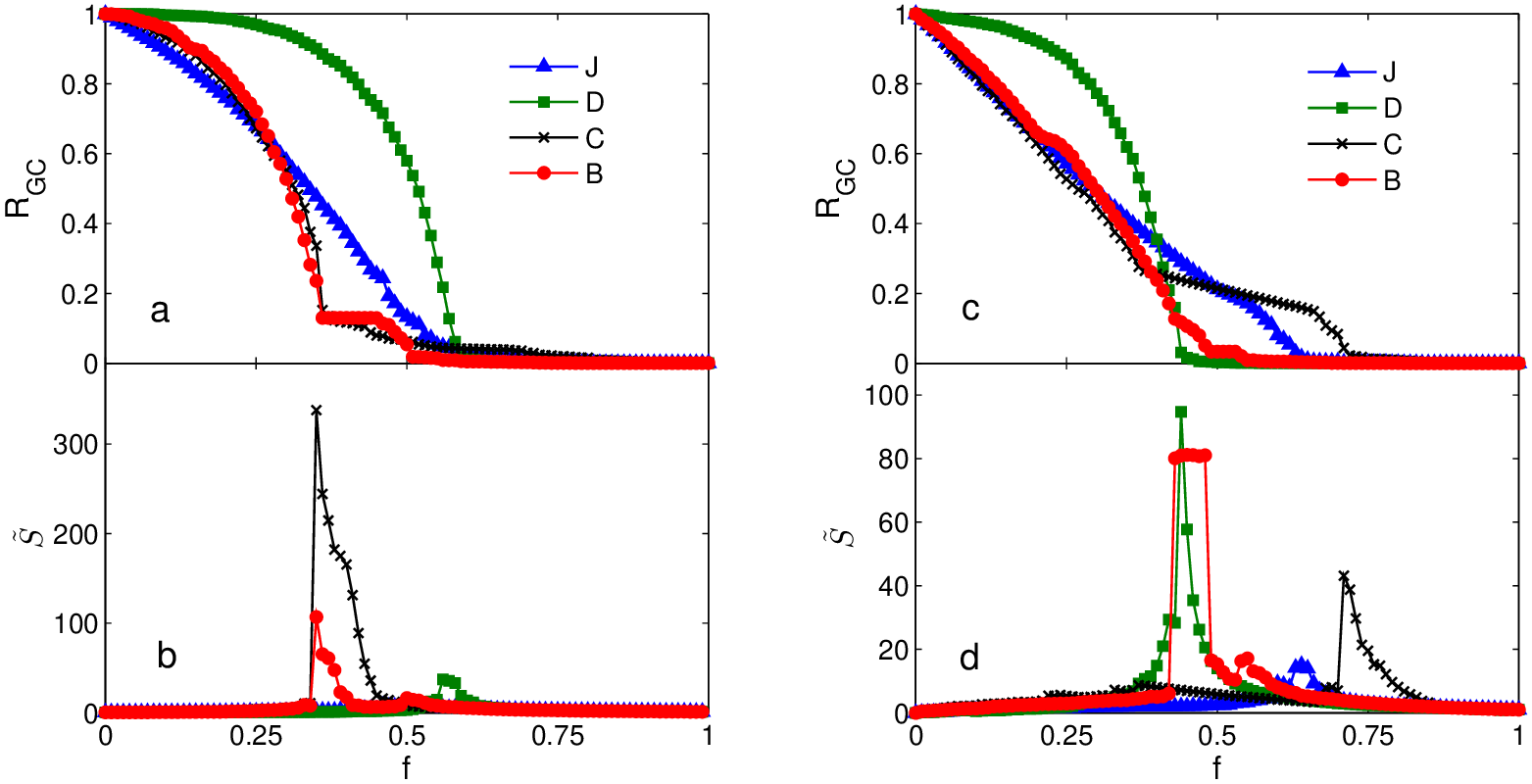}
\includegraphics[width=17cm]{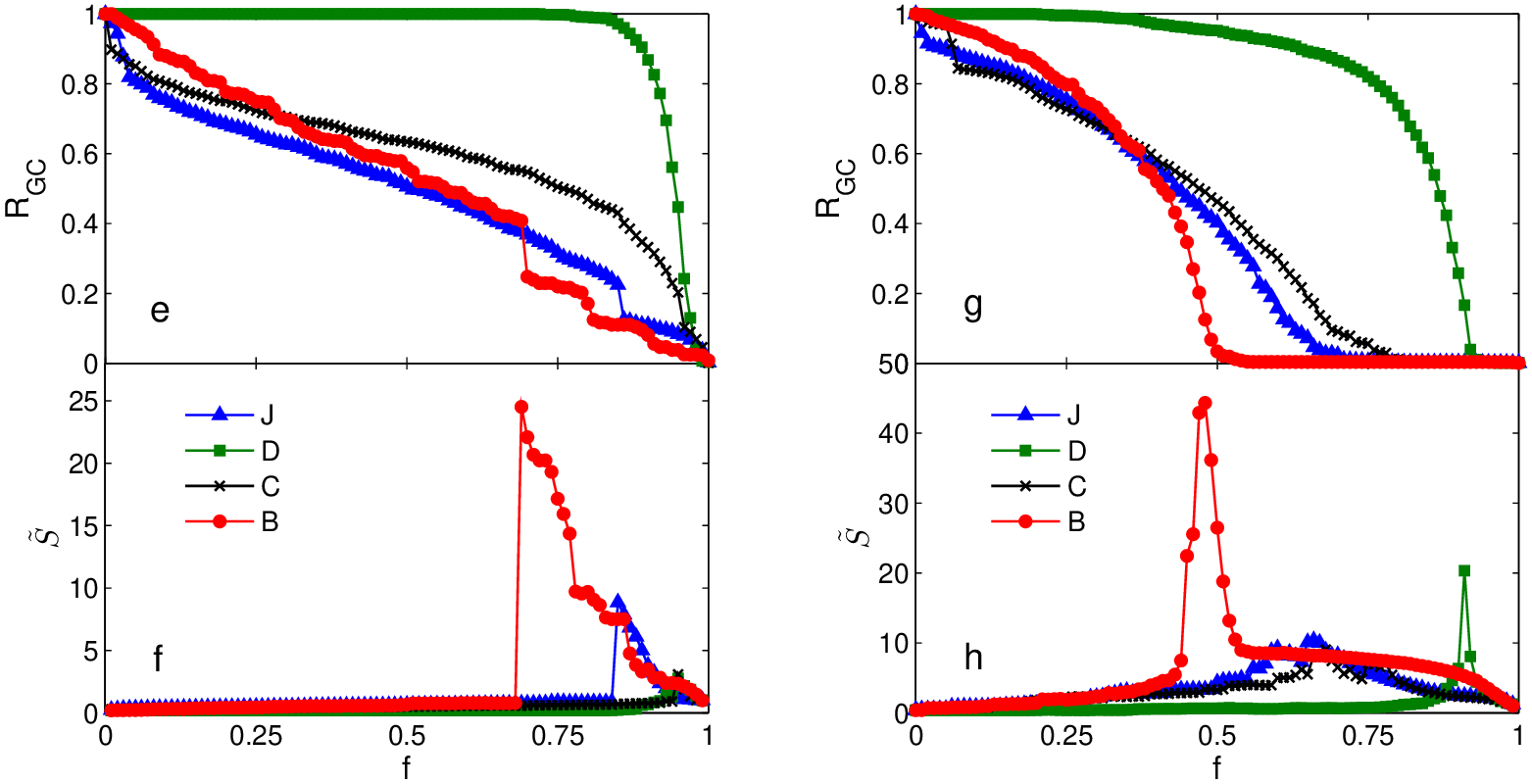}
\caption{\label{All_Result_on_Four} Edge percolation
results on four real networks. Plots (a) and (b) are for the PNAS
citation network, (c) and (d) for the social network of del.icio.us,
(e) and (f) for the political blog network, and (g) and (h) for the
astrophysics collaboration network. In each plot, four curves are
corresponding to the four structural indices, $J$ ($\triangle$),
$D$($\Box$), $C$($\times$), and $B$($\ocircle$). For the case of
Jaccard coefficient, the edges are removed in ascending order; while
for the other three cases, they are removed in descending order.}
\end{center}
\end{figure}

Empirical comparison is carried out on four networks. Besides the
PNAS citation network, the three others are the social network of
del.icio.us, political blog network and astrophysics
collaboration network. Delicious data was crawled in October
2009 from the famous online social bookmarking service del.icio.us,
where a user $i$ is considered to be a fan of another user $j$ if
$i$ has seen some bookmarks collected by $j$, and two users are
connected if they are fans of each other. The political blog network
is formed by weblogs on US politics and hyperlinks between them
\cite{Adamic2005}. The astrophysics collaboration network is
comprised of the coauthorships between scientists posting preprints
on the Astrophysics E-Print Archive from Jan 1, 1995 to December 31,
1999 \cite{Newman2001}. The basic statistics of these networks is
shown in Table~\ref{tab:Data_description}.

As shown in Fig. 4, for the PNAS citation network, edge betweenness
centrality and bridgeness perform best; for the social network in
del.icio.us, degree product and bridgeness perform best; for the
political blog network and the scientific collaboration network,
bridgeness performs best. Therefore, one can conclude that
bridgeness is remarkably better than the other three indices in
characterizing the edge significance in maintaining the global
connectivity. As a local index with light computational load,
bridgeness is expected to be applied in practice.

\section{Conclusion and Discussion}
In this article, we investigate the weak ties phenomenon in document
networks. Empirical analysis indicates that the weak ties, namely
the edges connecting less similar nodes in content, play more
significant role in maintaining the global connectivity. Inspired by
the strong correlation between the existence of an edge and the
strength of that edge \cite{Menczer2004,Cheng2009}, we believe the
edge significance in maintaining the global connectivity can be well
characterized by some indices depending solely on the topological
structure. Accordingly, we propose a local structural index, called
bridgeness. Compared with both the content similarity and three
well-known topological indices, bridgeness always leads an earlier
network disintegration in the edge percolation process, indicating
that bridgeness performs best in characterizing the edge
significance. This will help us in some real-life applications such
as controlling the spreading of diseases or rumor and withstanding
the targeted attacks on network infrastructures.

The percolation process is carried out on real-world networks and the
results illustrate that the networks are fragile when removing edges
with larger bridgeness. These networks are monopartite, i.e., nodes in
these networks are of no distinction. It will be necessary to generalize
the definition of bridgeness when facing other network configurations such
as bipartite networks, as Lind {\it et al} did in \cite{Lind2005}. Moreover,
empirical analysis in more large-scale networks will also be
a part of our future work.

\section*{Acknowledgements}
This work was partially supported by the National Natural Science
Foundation of China under grant numbers $60873245$ and $60933005$.
Z.-K.Z. acknowledges the Swiss National Science Foundation under
grant number 200020-121848. T.Z. acknowledges the National Natural
Science Foundation of China under grant numbers 10635040 and
90924011.

\end{document}